\documentclass{aastex}

\usepackage{emulateapj5,apjfonts}

\input{epsf}

\newcommand{\bez}{\begin{eqnarray*}}
\newcommand{\eez}{\end{eqnarray*}}
\newcommand{\be}{\begin{equation}}
\newcommand{\ee}{\end{equation}}
\newcommand{\beq}{\begin{eqnarray}}
\newcommand{\eeq}{\end{eqnarray}}
\newcommand{\bc}{\begin{center}}
\newcommand{\ec}{\end{center}}

\shortauthors{Stern et al.}
\shorttitle{Decline of GRBs and their luminosity function}

\begin{document}

\title{
The Decline of the Source Population of Gamma-Ray Bursts and Their
Luminosity Function
}

\author{B. E. Stern\altaffilmark{1,2,3}, 
Ya. Tikhomirova\altaffilmark{2,3}, 
R. Svensson\altaffilmark{3}} 

\altaffiltext{1}{Institute for Nuclear Research, Russian Academy of Sciences,
Moscow 117312, Russia}
\altaffiltext{2}{Astro Space Center of Lebedev Physical Institute,
Moscow, Profsoyuznaya 84/32, 117810, Russia}
\altaffiltext{3}{SCFAB, Stockholm Observatory, SE-106 91 Stockholm, Sweden}
 

\begin{abstract}
The source population of gamma-ray bursts (GRBs) declines towards the present
epoch  being consistent with the measured decline of the star formation 
rate. We show this using the brightness distribution of 3255 long BATSE GRBs 
found in an off-line scan of the BATSE continuous 1.024 s count rate records. 
The significance of this conclusion is enhanced by the detection of 
three GRBs with known redshifts brighter than 10$^{52}$ erg s$^{-1}$ 
during the last two years. This is  an argument in favor of the 
generally believed idea that GRBs are strongly correlated with 
the star production, at least on cosmological time scales, and 
favors the association of long GRBs with collapses of supermassive stars. 
However, we still cannot rule out neutron star mergers if the
typical delay time for binary system evolution is relatively short.
If we assume a steep decline of the GRB population at $z > 1.5$,
then their luminosity function can be clearly outlined. 
The luminosity function is close to a power law, $dN/dL \propto L^{-1.4}$, 
for low luminosities over at least 1.7 orders of magnitude. 
Then the luminosity function breaks to a steeper slope or to 
an exponential decline around $L \sim 3\cdot10^{51}$ erg s$^{-1}$ 
in the 50 - 300  keV range assuming isotropic emission.  
\end{abstract}


\section{Introduction}


In spite of the remarkable progress that has proved the cosmological origin
of GRBs, there remains a number of extremely important issues which are still 
not
resolved. Some of the main  issues  dealt with in this work
are the following.  (i) What is the 
the cosmological evolution of the source population of gamma-ray bursts (GRBs)?
Does it evolve as the star formation rate (hereafter the SF rate) or does it has
its  own specific evolution? (ii) What is the
luminosity distribution (or function)
of GRBs? (iii) What is the total rate of GRBs in the universe?

There are two competing approaches for such studies: 
(i) the ``statisitical'' one using large samples of poorly localized GRBs,
and (ii) the ``individual'' one, using the small sample of optically
followed-up GRBs, where we have additional information per event including the 
redshift and the intrinsic luminosity.     

The main data array for statistical studies was supplied by the Burst 
And Transient Source Experiment
(BATSE) (Fishman et al. 1989) onboard the {\it Compton Gamma-Ray Observatory} 
({\it CGRO}). The BATSE sample\footnote[4]{The
catalog is available at 
http://gammaray.msfc.nasa.gov/batse/grb/catalog/current/} 
is a few  times larger than the  yield of all other experiments that have 
detected
GRBs. It includes 2702 events in its final form.

Fitting the BATSE data to various
cosmological/evolutionary models has been the subject of many studies since the
start of the BATSE operation in 1991. For references to early works on
fitting the BATSE brightness distribution of GRBs to cosmological models,
see Bulik (1999). For a number of results setting an upper limit to the
width of the luminosity function of GRBs, see, e.g., Hakkila et
al. (1996).  Then Loredo \& Wasserman (1998) demonstrated that the 3rd
BATSE catalog does not constrain the luminosity function. Later
work using the larger sample of the 4th BATSE catalog gave few constraints.
Krumholtz, Thorsett, \& Harrison (1998) fitting the BATSE  sample at 
peak fluxes $P >$
0.42 photons s$^{-1}$  cm$^{-2}$
found that both a non-evolving GRB source population 
and SF evolutionary models fit the data even using 
standard candle GRBs. Wijers et al. (1998) demonstrated the same for 
the SF model.  
Totani (1999) showed that 
the SF model does not fit the data using the standard candle assumption.
Panchenko (1999) using the SF model accounting for the time delay due to binary
system evolution estimated the minimal width of the luminosity 
function as being two orders of magnitude in luminosity. Finally, the recent 
work 
of Porciany \& Madau (2001) deals with a larger sample 
including the BATSE non-triggered bursts of Kommers et al. (2000) extending 
the peak flux down to 0.18 photons s$^{-1}$ cm$^{-2}$. They found
that the data fit cannot  distinguish  between different variants of the GRB
source evolution at large redshifts. However, they did not check how sensitive 
the fit is to the evolution at {\it low} redshifts.

The general impression arising from fitting the BATSE data using cosmological
models was that this approach had little future. Indeed, the ``statistical''
approach demonstrated an agreement between the data and a wide set of models.
The very few constraints obtained were trivial. 

The main reason for the poor progress so far is the insufficient depth of 
the BATSE sample, i.e., the too 
narrow brightness range. The brightest burst has a peak photon flux of
160 photons s$^{-1}$ cm$^{-2}$. Bursts useful for a usual least $\chi^2$ 
fit are,
however,
at $P < 30$ photons s$^{-1}$ cm$^{-2}$. The BATSE trigger threshold is at 0.2
photons s$^{-1}$ cm$^{-2}$ but the bursts  near the threshold are
difficult to use because of poorly known threshold  effects. All works cited
above used the peak flux range above 0.4 or even above 1 photon s$^{-1}$
cm$^{-2}$, which is then narrower than two orders of magnitude. Loredo and
Wasserman (1998) showed that this is a too narrow range for obtaining 
constraints
from the fits. 

The ``individual'' approach gave a wealth of important data.
By itself this approach is, however, still unable to resolve the issues stated 
above.
Besides having a too poor statistics (presently we have only 17 GRBs with known
redshifts), the approach is subjected to very strong selection biases.
Nevertheless, this small sample  tells us that the luminosity function is at
least 2.5 orders of  magnitude wide and that it extends up to $\sim 3\cdot
10^{52}$ erg s$^{-1}$. 

After most previous work in this area was completed,
the following progress concerning the data accumulation has taken place:

 - BATSE obtained additional data until the deorbiting of {\it CGRO} in June
2000.
 
 - Searches for non-triggered bursts were performed by 
Schmidt (1999), Kommers et al. (2000), and Stern et al. (2000, 2001).
In the latter work, the statistics of useful GRB  events was increased by 
a factor 1.7 and the threshold effect was measured. Thus, the useful fitting 
range was extended down to 0.1 photons s$^{-1}$ cm$^{-2}$.

 - A sample of GRBs with known intrinsic luminosities (presently 17 events)
appeared due to the optical 
afterglow observations.

In this work, we use these advances including the
``individual'' GRB data. In addition, we include the 
brightest peak flux interval, which is statistically poor,  into the 
maximum likelihood fit and find that this peak flux interval is very
informative.   Thus we extend the fitting brightness range to 3 orders of
magnitude. This allow us to obtain a number of conclusive results.

In \S\S 2-4, we describe the set of fitted data, the fitting models including 
the cosmology, the source evolution, and the luminosity function 
of GRBs, and finally the
 the fitting procedure. In \S 5, we present results of the fits and show that  
the scenario of a non-evolving population of GRBs does not fit the data.
Instead,  we demonstrate that 
the GRB-population should decline approximately as fast as the star 
formation rate.
We also determine the approximate shape of the luminosity function
and give an estimate for the lower limit of the total rate of GRBs
as being 3000 GRBs per year in the visible universe (that is up to a 
reasonably large redshift).

\section{The data}


Probing various cosmological and evolutionary models  
we fitted the sample of 3255 BATSE GRBs longer than 1 s found by 
Stern et al. (2000, 2001) in the off-line scan of the BATSE continuous 
daily records in 1.024 s time resolution. This sample, which is selected from 
the catalog of Stern \& Tikhomirova\footnote[6]
{see http://www.astro.su.se/groups/head/grb\_archive.html},
is essentially uniform and has a corresponding efficiency matrix 
(measured by a test burst method), which is needed when 
fitting the weak end of the log $N$ - log $P$ distribution. (Hereafter
the term log $N$ - log $P$ distribution means the {\it differential} 
distribution of GRBs vs. the logarithm of the {\it peak photon flux}).
We excluded 
short bursts (consisting of one 1.024 s bin) from the analysis for two 
reasons: (i) short and long bursts could be separate phenomena, and 
(ii) the sample is incomplete regarding  short bursts as they have a 
lower detection efficiency and a wrong brightness estimate 
in 1.024 s time resolution. By excluding
one-bin events, we make our sample more homogeneous.

The brightness distribution of the GRBs in this sample was fitted using 
a hypothetical brightness distribution folded with the detection efficiency 
matrix described in Stern et al. (2001). 
This matrix was obtained using a sample of $\sim$ 11,000 
artificial test bursts, which were superimposed on the BATSE 
continuous records and then passed through the same procedure of search and
processing  as real GRBs. The efficiency matrix is approximately given by

\begin{equation}
F(c_e,c_m)= E(c_e)  \frac{1}{\sigma\sqrt{\pi}}
\exp \left[ -\frac{\log^2(c_m/c_{m\rm o})}{2\sigma^2} \right],
\end{equation}

\noindent
where $c_e$ is the expected and $c_m$ the measured count rate in units
of counts s$^{-1}$ cm$^{-2}$; 
$E(c_e)$ = $0.70(1-\exp[-(c_e/c_{e\rm o})^2])^\nu$
is the efficiency function
with fitted parameters $c_{e\rm o}$ = 0.097 counts s$^{-1}$ cm$^{-2}$, 
$\nu$ = 2.34; the  log-normal
factor  describes the relative error of the measured  count rate, $\sigma$ =
$0.09  (0.08/c_e)^{1/2}$,  and the  selection  bias is crudely  expressed as
$c_{m\rm o}$ = $c_e+0.05\exp (-c_e/0.05)$.

In order to constrain the intrinsic luminosity function
(hereafter the luminosity function or the LF), we used the sample 
of gamma-ray bursts with measured redshifts\footnote[7]
{see, e.g., http://www.aip.de/~jcg/grb.html}.  We cannot
infer the LF from this sample as it is subjected to strong selection
biases. This is demonstrated in section 3.3. 
The redshift data, however, give us a useful piece of information, i.e., the
existence of very   intrinsically bright GRBs. 
Three of the intrinsically brigthest bursts are:
GRB990123, GRB991216, and GRB000131 (named by dates), with redshifts 1.6
 (Djorgovski et al. 1999),
1.02 (Vreeswijk et al. 1999), and 4.5 (Andersen et al. 2000), 
respectively, and with BATSE  peak fluxes 
16.4, 67.5, and 6.3 photons s$^{-1}$ cm$^{-2}$  in the 50 - 300 keV range, 
respectively (estimated using the BATSE catalog). If they were emitted at 
$z=1$, their peak fluxes taking into account the ``K-correction''
(i.e., the
correction due to the spectral redshift  effects on a fixed spectral band of the
detector)  would 
be 45, 69, and 84 photons s$^{-1}$ cm$^{-2}$, respectively, assuming
the cosmological parameters ($\Omega_{\rm M}, \Omega_\Lambda) = (1,0)$.  
Hereafter, we use
the photon peak flux at redshift $z$=1, $I$, as a measure of the intrinsic
brightness. We, furthermore, use the intrinsic brightness interval of these 
three events, 
$I > 40$ photons s$^{-1}$ cm$^{-2}$, to constrain the LF when fitting 
the BATSE log $N$ - log $P$ distribution.   The choice of the three 
brightest events for this purpose is
somewhat arbitrary. We cannot use a much wider brightness interval because 
of a brightness dependent selection bias. On the other hand, we can neglect 
such problems for the narrow
brightness range of these three events.
The data we fitted are presented in
Figure 1.

 In order to impose a proper constraint on the LF we must 
estimate the sampling function for strong GRBs. With the 
sampling function we mean the probability that a burst will be detected, 
localized, its afterglow observed and its redshift measured.
This function evolves with time. It was zero before 1997. Then this function 
was limited by the field-of-view of the two
Beppo-Sax Wide Field Cameras,
$\sim 0.08$ of the sky\footnote[8]{see http://www.asdc.asi.it/bepposax/}, 
as this was the main instrument supplying precise
coordinates of GRBs   during 1997 - 1998.


In 1999 and 2000, many precise localizations were made by  other systems,
with most of them being made by
the interplanetary network {\it Ulysses}/{\it Konus}/{\it Near} 
(see the IPN home page\footnote[9]{http://ssl.berkeley.edu/ipn3/index.html}).
This means 
that in this period the sampling function became larger and for very bright 
events it could, in principle, approach unity as all instruments of the IPN had 
a 4$\pi$ field-of-view. Actually it should be considerably less as
in the same period, three very strong BATSE events (triggers 7301, 7491, 
7595) were not localized. For one strong BATSE event (trigger 7954, GRB000115) 
an
X-ray transient was found, but no optical transient. With this background, let 
us 
take a conservatively high estimate of the sampling function, $S_{40}=0.5$
for GRBs  with $I >$ 40  photons s$^{-1}$ cm$^{-2}$, 
and the conservatively low estimate 
of the rate of these GRBs, $N_{40} = 3$ yr$^{-1}$, in the visible universe.

\section{Fitting Models}

\subsection{Cosmology}
We tried two sets of cosmological parameters, the flat 
matter-dominated universe, commonly used in most of previous works: 
($\Omega_{\rm M}, \Omega_\Lambda$) = (1,0) and the vacuum-dominated cosmology, 
which is supported by recent data 
($\Omega_{\rm M}, \Omega_\Lambda$) = (0.3, 0.7) (see, e.g., Lukash, 2000).
Hereafter, these two models are denoted as M-models and $\Lambda$-models,
respectively.  
The distribution of GRBs over redshift for a non-evolving (NE)
population for $\Omega_{\rm M} + \Omega_\Lambda = 1$ is

\begin{equation}
\frac{dN}{dz} \propto \frac{1}{1+z}
\frac{1}{\sqrt{\Omega_\Lambda+\Omega_{\rm M}(1+z)^3}}
\left(\int_0^z \frac{dz'}{\sqrt{\Omega_\Lambda+\Omega_{\rm M}(1+z')^3}}\right)^2
\end{equation}

\noindent
The photon number ``luminosity'' distance is defined by 

\begin{equation}
d_{\rm L} = {c\over H_0}\sqrt{1+z}
\int_0^z \frac{dz'}{\sqrt{\Omega_\Lambda+\Omega_{\rm M}(1+z')^3}},
\end{equation}

\noindent 
where $H_0$ is the Hubble constant, which is assumed to be 75 km s$^{-1}$
Mpc$^{-1}$, when estimating  the luminosities of GRBs. For the standard 
luminosity distance, the factor $\sqrt{1+z}$ is replaced with $1+z$. 
 
\subsection{Evolution of the Source Population}

Another important component of the model is the evolution of the population
of GRB sources. We probed four cases: a non-evolving population and 
three
evolution functions correlated with the history of star formation 
following Porciani \& Madau (2001). The declineing phase of the star formation
(SF) rate
at $z <$ 1.5  is a relatively well measured function of $z$.  
Its history at $z >$ 2 is, however, controversial. This issue is discussed in 
Porciani \& Madau (2001) giving the  relevant references.
Below we reproduce three versions of the SF evolution suggested in that work: 

\begin{equation}
R_{\rm{SF1}}(z) = \frac {0.3e^{3.4z}}{(e^{3.8z}+45)} \quad 
{\rm M_{\odot}} \ {\rm yr}^{-1} {\rm Mpc}^{-3} \ ,
\end{equation}
i.e., decreasing SF at $z > 1.5$,

\begin{equation}
R_{\rm{SF2}}(z) = \frac {0.15e^{3.4z}} {(e^{3.4z}+22)}  \quad
{\rm M_{\odot}} \ {\rm yr}^{-1} {\rm Mpc}^{-3} \ ,
\end{equation}
i.e., roughly constant SF at $z > 2$,

\begin{equation}
R_{\rm{SF3}}(z) = \frac {0.134e^{3.05z}}{(e^{2.93z}+15)} \quad
{\rm M_{\odot}} \ {\rm yr}^{-1} {\rm Mpc}^{-3} \ ,
\end{equation}
i.e., increasing SF at large $z$.

\noindent
Models will from now on be denoted as NE,M; SF1,$\Lambda$; and so on,
where M and $\Lambda$ denotes the two types of cosmologies.

The next step is the generation of the standard candle log $N$ - log $P$ 
distributions. At this step, we introduced an additional 
broadening of the  observed brightness distribution due to the 
K-correction which depends on the type of GRB spectrum.
For this purpose, we obtained the log $N$ - log $P$ distributions 
using Monte-Carlo simulations. To each simulated GRB we prescribed  one of 54
spectra of bright  BATSE bursts  parametrized by the Band expression
(Band et al. 1993). All these template spectra were assumed to be emitted at 
$z$= 1. Then we sampled the $z$ of the burst and the corresponding  
K-correction for the 50 - 300 keV band was applied to the apparent brightness 
of the simulated GRB. 

The resulting log $N$ - log $P$ distributions are shown in Figure 2.
 If we neglect the K-correction which depends on the GRB spectrum, then 
each distribution in Figure 2 is a direct reflection of the
corresponding redshift distributions of GRBs.
The standard candle luminosity (before applying the K-correction) corresponds to
a peak photon flux of 1 photon s$^{-1}$ cm$^{-2}$  in the 50 - 300 keV  band at
$z$=1.

If we vary this value  when fitting the data, none of the 8 models
(two cosmological cases, four evolutionary cases) would still
fit the observed log $N$ - log $P$ distribution of 3255 long BATSE GRBs
(shown by the  crosses in Figure 2).  The fact that the evolution of 
the SF1 type
with the standard candle luminosity function cannot fit data was shown by
Totani (1999) and Lloyd \& Petrosian (1999).  
 The NE model gives the smallest deviation in this case, 
but still the value of $\chi^2$ is unacceptable (61 for 
27 degrees of freedom). However, as it will be shown below, the discrepancy 
in the case of the NE model can not be compensated by any hypothesis of the 
luminosity function.
 
The slopes of the three log $N$ - log $P$ distributions for an evolving SF rate 
are
close to the  Euclidean 
slope -3/2  starting from a peak flux $P$ $\sim 1$ photon s$^{-1}$ cm$^{-2}$,
which at this standard candle brightness corresponds to $z=1$. 
The agreement with the Euclidean slope is an  accidental
coincidence due to the superposition of cosmological and  evolutionary effects.

\subsection{Parametrization of the Luminosity Function}

 As was stated above, the luminosity distribution of events with known 
$z$ cannot be used as a base for the LF model in the whole brightness 
range. This fact is clear from Figure 3 where we present log $N$ - log $P$ 
distributions that would 
give the sample of 17 GRBs with known absolute luminosities for the NE, SF2
and SF3 evolutionary models. All models give a striking disagreement with 
the data.
The only way to reduce this disagreement is to assume an unreasonably sharp 
increase of GRBs at large redshifts which, in turn, will contradict the 
redshift data. Therefore the shape of a hypothetical 
LF remains arbitrary (except the constraints imposed on the brightest end of 
the LF).

In order to get a handle on the LF of GRBs, we tried  
different types of functions that describe common shapes of wide distributions
in nature: the log-normal distribution (LGN), a truncated power law (TPL), 
a power law with an exponential cutoff (PLexp), and a broken power law (BPL). 

LGN: $dN/dI = C\cdot \exp(-{(\ln^2(I/I_0)/2\delta^2})$, with three
 free parameters $I_0$, $\delta$, and $C$.

TPL: $dN/dI = C\cdot I^{\alpha -1}$ for $I_1 < I < I_0$ and 0 outside this 
interval. Free parameters  are $\alpha$, $I_1, I_0$, and $C$. In some fits, 
$I_1$  was fixed to $-\infty$  leaving 3 free parameters.

PLexp: $dN/dI = C\cdot I^{\alpha-1} \cdot \exp(-I/I_0)$, with three free 
paremeters $\alpha$, $I_0$, and $C$.

BPL:  $dN/dI = C\cdot I^{\alpha -1}$ for $I_1 < I < I_0$, $dN/dI = 
C_1\cdot I^{\alpha +\beta -1}$ for $I_0 < I < I_2$ and $dN/dI = 0 $ outside
the [$I_1,  I_2$] interval. Free parameters are
$\alpha$, $\beta$, $I_1$, $I_0$, and $C$, while $I_2$ was fixed to a value
above the  maximum observed GRB brightness. 

 We also considered a smoothed version of the broken power law:

SBPL: $dN/dI=C I^{\alpha - 1}/(1+(I_0/I)^{_\beta})$

For technical convenience, we measure the intrinsic brightness as
peak count rate or peak photon flux, $I$, in the 50 - 300 keV range produced by 
a 
GRB at a  distance 
corresponding to $z=1$. The absolute peak luminosity of the GRBs  is related  
to $I$ as $L = I\cdot 3\cdot 10^{50}$ erg s$^{-1}$ assuming isotropic emission.
Below we present the main results on the LF both in $I$ and in 
absolute luminosity units. 

\bigskip
\section{The fitting procedure}

We used the forward folding 
method when fitting the observed distribution of GRBs,
i.e.,  the hypothetical brightness
distribution was convolved with the  efficiency matrix (1) and fitted to the
observed distribution of GRBs over peak count rate (crosses in
Figure 2).  This distribution was represented by 29 data points spaced by 0.1 in 
log $P$ in the interval 0.067 - 50 photons s$^{-1}$ cm$^{-2}$ in the 50 - 300
keV range.
During 9.1 years, there are  3 GRBs detected by BATSE that are brighter  
 (the rightmost cross in Figure 2  and one GRB at 
160 photons s$^{-1}$cm$^{-2}$,
which is not shown). We  treat this brightness range separately,
estimating the likelihood function of the fit for each peak flux interval. For 
the
main interval, this is  the standard $\chi^2$ probability function. For the tail
of the brightness distribution, the likelihood is the Poisson probability to
sample not more then 3 events  brighter than 50 photons s$^{-1}$ cm$^{-2}$  at 
the
given number of such events for the full observation period, 
$M_{50}$, predicted by the model.  The final
likelihood finction is the product of these two factors.

Instead of the usual $\chi^2$ minimization procedure, we explore the 
parameter space 
sampling $\sim 10^5$  random points. This is sufficient to find the minimum
of $\chi^2$ with a good accuracy while at the same time investigating the 
$\chi^2$ ``topography''  of the parameter space region. This method  does not 
work
if the ``valley'' of the minimum has a very small volume in parameter
space. For our cases, the minima are wide and smooth enough.

 The maximum likelihood point for the whole sample of points in parameter space
represents the {\it unconstrained} fit, where the requirement of the redshift
data,
$N_{40} > 3$ yr$^{-1}$, was ignored. The subsample selected including this 
requirement
represents the {\it constrained} fit. The effective number of degrees of freedom
(DOF) in the  second case is smaller by unity as compared to the 
unconstrained fit. We present results  of both the constrained and
the unconstrained fits in order to demonstrate the role  of the redshift data 
and 
the possible effects of the uncertainty in the estimate of the rate of 
intrinsically strong GRBs.

 The best fit parameters are presented in Tables 1 and 2. Table 1
includes fits using various LF  models and selected cosmological models.
Table 2 summarizes  fits using a broken power law LF for all cosmological 
models and provides data to compare their relevance using a Bayesian approach.

\section{Results}

\subsection{Rejection of models without source evolution}

 The best unconstrained fit using the NE models has $\chi^2$ = 36.8 for 25 
degrees 
of freedom (Table 1, truncated power law, TPL), which is marginally 
acceptable if we ignore the tail of the 
brightness distribution. For the tail, the model predicts $M_{50}$ = 11.5,
   while the real number is 3 yr$^{-1}$ and the corresponding maximum
likelihood  drops down to 1.3$\cdot 10^{-4}$  (see Table 1).
When we impose the redshift data constraint, $N_{40} > 3$ yr$^{-1}$, the maximum 
likelihood factor that we can obtain using the NE,$\Lambda$ model is  
3.2$\cdot 10^{-6}$, which is for the case of a broken power law LF (see Table 
2).
For the NE,M model, the results  are even worse.

 The likelihood factor is not yet a rejection factor for the NE model, because 
we 
cannot exclude some bias or contamination which would increase the $\chi^2$. To
reject  the NE hypothesis, we should demonstrate a good fit for other equally
simple and  reasonable models. 
Indeed, data fits with the SF models are much better (see Tables 1 and 2). 
Their maximum likelihood factor is about 0.02. This is the case when 
we can apply the Bayesian approach. The estimate of the rejection level
for models with nonevolving GRB source population is the ratio
of the maximum NE likelihood factor (NE,$\Lambda$ model in Table 2) to that for 
SF models (e.g, SF1,$\Lambda$ in Table 2),  which is
$1.4\cdot 10^{-4}$.

  Figures 4, 5, and 6 demonstrate the differences between fits
using NE and SF models. As seen in Figure 5,
the high brightness slope of the log $N$ - log $P$ distribution for
the broken power law NE model is too flat and it cannot be made considerably
steeper by modifying the LF. Note that the NE standard candle
log $N$ - log $P$ distribution with $I$ =1 photon s$^{-1}$ cm$^{-2}$  is already 
flatter than the observed  log $N$ - log $P$ distribution. Then if the LF is
extended  to
$\sim$ 100 photons s$^{-1}$ cm$^{-2}$, the model tail will be considerably 
flatter
than the observed tail independently of how the LF is extended. 
  
 Looking at the integral distribution of real GRBs in Figure 6, one notices
that it declines faster than an Euclidean distribution. However, this is still
not a  statistically significant fact. The probability of such a deviation from
the -3/2 slope by chance is 0.1, which is  relatively large (integral
distributions are  known to produce an illusion of statistically significant
features from  fluctuations). Most probably we are just dealing with a
moderate fluctuation.  Nevertheless, the observed slope could really be steeper
than the  Euclidean one if the decline of GRBs is steep enough. More data   
are required to clarify this issue. 

 The rejection of the NE models is significant even without the redshift data.
The unconstrained rejection factor, taken as the ratio of the unconstrained 
likelihoods of the NE,$\Lambda$ and SF1,$\Lambda$ models
 is 
$\sim 2\cdot 10^{-3}$ (see Table 2).
This means that the result is not very sensitive to the value of the 
constraining
$N_{40}$. If we, e.g., overestimate the rate of intrinsically strong GRBs by a
factor 3 (suppose that it is a fluctuation) and that
$N_{40} = 1$, then the rejection factor is 4$\cdot 10^{-4}$.

While the NE models are rejected using the low redshift behavior, one does not 
obtain 
any  preference for  a certain kind of SF evolution at large redshifts.
Different SF models give similar likelihood results (see the Tables)
except for the SF3 scenario, which gives a slightly worse fit. 
Furthermore, the data do not allow us to 
distinguish  between matter-dominated and vacuum-dominated cosmologies.

 \subsection{ The Shape of the Luminosity Function  }

 Once the NE models have been rejected at a significant level, we now
concentrate on the SF models, i.e., models  with evolution of the GRB source
population.  There 
are two clear features of the LF that we see for all SF
models: a near power law  interval at the lower brightness range and a break or
turnover towards the  bright end of the distribution. 
Attempting to replace this construction of the LF by a log-normal LF 
gives a decrease of the maximum likelihood by 2 orders of magnitude (see,
e.g., the two SF1 models in Table 1).

If we study the $\chi^2$ topography for the broken power law LF, we find
a power law fragment at least 1.7 orders of magnitudes wide 
and there is no upper limit on its width. It can be arbitrarily extended to 
lower brightnesses  (see Fig. 7).  Indeed, one can see from Figure 7 
that the $\chi^2$ distribution reaches its 
asympthotics: by extending the LF further to lower brightnesses does not 
affect the model in the range of the data points.

 On the other hand, the slope of this power law is surprisingly well 
constrained, especially for the SF1 model (Fig. 7), and it is sligtly sensitive 
to
the  choosen SF behavior at
large redshifts. The shape of the $\chi^2$ minimum changes, however. Larger SF
rates at high redshifts allow flatter slopes $\alpha$ (see Fig. 8). 
The $\chi^2-\alpha$ plots for a matter-dominated cosmology, which are not 
shown,
are very close to those shown in Figure 8. They are just slightly narrower and 
more symmetric.

 A break or a turnover is necessary at a high significance level. Its removal 
increases
$\chi^2$ by $\Delta \chi^2 \sim 30$ for SF1 and $\Delta \chi^2 \sim$ 19 - 25 for 
SF2
models  (see Table 2).
The properties of the break are, however, less certain than the parameters
of the  power law fragment. The  $\chi^2$ minimum is non-parabolic, asymmetric
and relatively wide as seen in Figure 9. All we can say is that some
turnover in the power law LF is required at an intrinsic brightness 
of about 10 photons s$^{-1}$ cm$^{-2}$  at $z=1$, or $\sim 3\cdot
10^{51}$ erg s$^{-1}$ for isotropic emission. The fitted position of the break 
is
slightly sensitive to  the cosmological model but the difference is within
the statistical errors.
We, however, can not
distinguish between a power law break and an exponential cutoff of the LF.
Both give a good $\chi^2$ and the same maximum likelihood  
(compare Tables 1 and 2). Summarizing, we can claim that the behavior
of the LF below $I \sim 3$ photons s$^{-1}$cm$^{-2}$ or $10^{51}$erg s$^{-1}$ 
(in the 50 - 300 keV range) is close to the power 
law $dN/dI \propto I^{-1.5}$ (or, more exactly, the power law index $\alpha-1$ 
can
vary  from  $\sim -1.35$ to $\sim -1.5$ depending on the model). Then it 
breaks to a steeper slope. A smooth break (i.e., the parametrization SBPL in
section 3.3) also gives a good result. The likelihood is 0.016 for 
the SF2,$\Lambda$ model when the difference in the slopes is large enough,
i.e., $\beta < -3$.

Figure 10 shows a set of the best fit LFs for different models.
The reason why the best fit LFs is more or less well defined is clear when
comparing  the LF curves with the BATSE log $N$ - log $P$ distribution shown by
the crosses.   The latter should be a
convolution of the former with the standard candle curves in Figure 2.
For the SF1 and SF2  models, the  standard  candle curves are ``narrower'' then
the BATSE log $N$ - log $P$ distribution. Therefore the LF should roughly
correspond to the main features  of the observed log $N$ - log $P$ distribution: 
a
power law with a turnover. In the case of the SF1 model with narrower
redshift distribution,  the LF is closer to the observed brightness 
distribution (see Fig.10).
 It is  natural  that the required turnover of the LF
is sharper than that of the log $N$ - log $P$ distribution with the sharp
LF turnover being smoothed by the convolution.

\subsection{The lower limit on the total GRB rate}

 The last column of Table 2 shows the lower limits on the total rate of GRBs
in the visible universe. These limits correspond to $\Delta\chi^2 = 7.$,
i.e., to $1 \sigma$ of a $\chi^2$ distribution with 25 DOF.
Note that their values  are obtained using an abrupt cutoff of the LF 
at the dim end. 
For the SF2,$\Lambda$ model, 
the highest allowed cut off is at $\sim 0.4 \cdot 10^{50}$ erg s$^{-1}$.
 A more realistic smoother cut off would
give a higher lower limit. We suggest that the value
3000 GRBs per year gives a realistic estimate of the minimum GRB rate.
This estimate is for {\it long} GRBs only.
The result depends on the SF model in a natural way predicting a larger 
result for a higher SF rate at large redshifts.

 For the SF2,$\Lambda$ model of the GRB source evolution, the highest minimum 
comoving GRB rate at large redshifts is
$\sim 3$ yr$^{-1}$ Gpc$^{-3}$.
At the present epoch, it becomes 0.13 yr$^{-1}$ Gpc$^{-3}$. This lower
limit coincides with the value claimed by Porciani \& Madau (2001)
as the estimate derived from a fit of the data of Kommers et al. (2000). 
It is, however, very difficult to compare results as the slopes and the
ranges of the LF are different. In fact, the shape of the LF derived 
by Porciany and Madau has the same parametrization as one of
our models, a power law with an exponential cut off, but, neverheless, 
contradict our results predicting a different slope and a finite
estimate for the low brightness cutoff of the LF. The latter could be
a consequence of the different behavior of the log $N$ - log $P$
distribution near the threshold in Kommers et al. (2000).     
  
  Our estimate is close to the result of Schmidt (2000) which is 
$\sim$ 0.2 yr$^{-1}$ 
Gpc$^{-3}$ at  $z=0$  derived assuming that the present GRB 
rate is 10 times less than at $z\sim 1.5$ (in our case the decline is a 
factor 23). On the basis of this estimate, we can discuss the possible 
association of GRB980425 and the supernova SN 1998bw 
(Bloom et al., 1999, see also Lamb, 1999 for a discussion). 
Schmidt (2000) estimates the probability 
of events such as GRB980425 if the association is real as being as low as
$10^{-3}$ yr$^{-1}$.  Indeed, the distance to SN 1998bw is $\sim$ 40 Mpc. The
corresponding  sampling volume is 2.7$\cdot 10^{-5}$ Gps$^3$. Our estimate of
the GRBs rate  0.13 yr$^{-1}$ Gpc$^{-3}$ with the cutoff $\sim 3 \cdot 10^{49}$
erg s$^{-1}$. The 50 - 300 keV peak luminosity of GRB980425 is 3$\cdot
10^{46}$ erg s$^{-1}$ (our rough estimate). If the luminosity function extends
down to this  luminosity with the determined slope $\alpha - 1 \approx -1.4$, 
then
the rate of GRBs  above this luminosity will be $\sim 20$ times larger, i.e., 
2.6  yr$^{-1}$ Gpc$^{-3}$ at $z=0$ and a corresponding rate of 
0.7$\cdot 10^{-4}$ yr$^{-1}$ in the sampling volume. 
Furthermore, one should take into account
the probability that such events will be localized with the accuracy of a few 
arcseconds which reduces the estimate by at least an extra order of 
magnitude. Thus
the probabliity of occurance and good localization of an event such as
GRB980425 within  $\sim 40$ Mpc with the power law behavior of LF down to 
$10^{46}$ erg s$^{-1}$ is below $10^{-5}$ per year. Such a small probability 
can hardly be compensated by a break in the LF below the observational
cut off.
The break must be so steep that it will 
affect the observed log $N$ - log $P$ distribution near the BATSE threshold.
This is a strong argument that GRB980425 might represent a different 
phenomenon
that does not overlap with classic GRBs in its luminosity function (but at 
the same time being 
indistinguishable from typical GRBs in its general appearence). A more
probable possibility is that we are dealing with an accidental coincidence.

\section{Conclusions}

 The source population of GRBs sharply declines from $z \sim 1.5$ towards 
the present epoch. This fact is established at the confidence level 
$10^{-4}$. The measured decline of the star formation rate being used as the
evolution hypothesis for the source population of GRBs fits the data 
satisfactorily.

 This is what is expected according to prevailing view of GRBs
as being the  product of stellar evolution.  Such a scenario was already 
successfully applied for the description of the observed log $N$ - log $P$ 
evolution, e.g., by Wijers et al. (1998), Panchenko (1999), Kommers et al. 
(2000), Schmidt (2000), and Porciani \& Madau (2001). However, nobody has 
quantitatively demonstrated that an evolution similar to the SF rate at some 
range of $z$ is 
{\it necessary} to describe data. We can mention just the work of Schmidt
(2000) where the SF rate hypothesis fits the bright tail of the log $N$ - log
$P$ distribution better (by a visual impression) than the NE hypothesis.
However,  no quantitative comparison of different scenarios has been done. 
We have here  demonstrated that the decline of the GRB population consistent
with the SF decline is a {\it necessary} requirement.

The main issue is whether the GRBs 
are 
associated with the collapse of massive stars (collapsars) or
the merging of neutron stars (mergers). On the small scale, these scenarios 
differ regarding the expected correlations of GRBs with star formation regions 
in
galaxies: collapsar GRBs should be well correlated, merger GRBs should show no
correlation with the star formation. This fact stimulated searches for such
correlations using optical GRB afterglows (see, e.g., the review of
van Paradijs, Koveliotou, \& Wijers, 2000).  
On the cosmological scale, such correlations should 
exist in both cases. However, in the merger scenario the occurance of GRBs 
will be described
by the SF rate convolved with a delay function. If the latter extends to 
a few billions years, then the decline of the GRB population will be 
considerably flatter. For estimates of the delay function for 
mergers, see Portegies Zwart \& Yungelson (1998) and Panchenko et al. (1999).

Qualitatively, our results favor the collapsar scenario.
 A very interesting situation occurs if the very steep, steeper than -3/2, 
slope of the 
log $N$ - log $P$ tail as shown in Figure 5 will persist to larger $P$
({\it Ulysses} data reduced by Atteia, Boer \& Hurley, 1999 
indicate that this can be the case).  Then one will have to accept that the GRB
progenitors are very massive  collapsars whose
population can decline faster than the general SF rate. 

But, at present,
we do not have sufficient statistical arguments to rule 
out mergers. 
We believe that the data can give tight constraints 
on the
delay function  for the merger scenario and the latter can be challenged by 
such constraints. However, we leave such estimates for future studies for 
the reason that in order to obtain solid conclusions, it is worth to
incorporate the  {\it Ulysses} data, which would provide 
at least a doubling of the statistics of the brightest GRBs.

 Our results
concerning the luminosity function of GRBs confirm the conclusion of 
Loredo \& Wasserman (1998) that the width of 
the luminosity function is not constrained by the BATSE data being wider than 
two orders of magnitude.  
 The shape of the LF tried in the most of previous works 
was a truncated power law or a lognormal distribution which satisfied earlier 
data in a narrower brightness range. Schmidt (2000)
used a broken power law hypothesis similar to one of our models.
The position of the break is consistent with our results.

 The interpretation of the shape of the  
luminosity function is beyond the scope of this work. 
 In principle, such type of distributions - broken
or exponentially cut power laws are common in nature. Then the break 
implies some physical limit such as a finite energy source.
We believe that the outlined shape of the LF
might be a useful clue in the development of physical models for the GRB
emission.


This research made use of data obtained through the HEASARC Online Service
provided by NASA/GSFC.
This work was supported by the Swedish Natural Science Research Council,
the Royal Swedish Academy of Science, the Wenner-Gren Foundation for 
Scientific Research, and the Russian Foundation for Basic Research grant
00-02-16135.  We thank the anonymous referee for useful suggestions.



\begin{deluxetable}{llllllll}
\tabletypesize{\scriptsize}
\tablewidth{0cm}
\tablecaption{Best  fits of data to some cosmological models with different 
forms of the LF.}
\tablehead{
\colhead{Model}
&\colhead{LF}
&\colhead{$\chi^2$}
&\colhead{$M_{50}$}
&\colhead{Likelihood}
&\colhead{$\alpha$}
&\colhead{$\delta$}
&\colhead{$\Delta I_0$}
}
\startdata
NE,$\Lambda$&TPL&36.8,39.8&11.5&1.3$\cdot$10$^{-4}$
&-0.37$^{+0.27}_{-0.31}$& & 2.8, 10.7 \\
NE,$\Lambda$&LGN&44.3,42.5&16.6&1.2$\cdot$10$^{-6}$&  &1.9&0.032, 
0.36 
\\
NE,$\Lambda$&PLexp&54.0, 49.7
&18.7&2$\cdot$10$^{-8}$&-0.21$^{+0.71}_{-0.38}$& &15, 21 \\
SF1,$\Lambda$&LGN&44.5,39.8&6.0&$3\cdot10^{-4}$&&2.0 &0.032, 0.12\\
SF1,$\Lambda$&PLexp&31.2,30.7&7.0&0.022&-0.43$\pm$0.09 &&14, 28  \\
SF2,$\Lambda$&PLexp&34.2,31.7&6.7&0.016&-0.36$^{+0.13}_{-0.10}$ &&11, 
 33  \\
SF2,M&PLexp&32.7,31.4&6.9&0.018&-0.36$^{+0.12}_{-0.08}$ && 12, 26 
\\
\enddata
\label{fits}
\tablecomments{
For models and LF parametrizations, see \S\S 3.1-3.3.  The fit with the LF
unconstrained by the redshift data is given only for TPL case. For all other 
given
fits,  the LF is constrained by $N_{40}  > 3$ yr$^{-1}$.
The $\chi^2$ values are given at the maximum
likelihood and the minimum $\chi^2$ points, respectively.
The ''Likelihood'' is the product of the $\chi^2$ probability and 
the Poisson probability of sampling not more than 3 events at a given 
expected number $M_{50}$. $M_{50}$ is given at maximum likelihood point.
$\alpha$ is the slope of the
power law part of the LF. $\delta$ is the width of the log-normal
distribution (see \S 3.3). $I_0$ is the upper cut off brightness of the 
truncated
power law LF,  the center of the log-normal distribution,  or the
exponential cutoff energy for the PLexp LF (see
\S 3.3).  The errors in
$\alpha$ and the confidence interval, $\Delta I_0$, for $I_0$ correspond to
$\Delta \chi^2 = 4$ (corresponding to $2\sigma$ in the conventional 
interpretation. 
However, one should be careful with such an interpretation when the results
depends on missing data points below the threshold). 
}
\end{deluxetable}

\begin{deluxetable}{lllllllll}
\tabletypesize{\scriptsize}
\tablewidth{0cm}
\tablecaption{Best fits of data to all evolution models with a broken power law 
LF}
\tablehead{
\colhead{Model}
&\colhead{$\chi^2 $}
&\colhead{$M_{50}$}
&\colhead{Likelihood}
&\colhead{Uncon. lkh.}
&\colhead{$\alpha$}
&\colhead{$\Delta I_0$}
&\colhead{$\Delta \chi^2$}
&\colhead{$N_{\rm tot}$ yr$^{-1}$}
}
\startdata
NE,M&39.1&16.8
&6$\cdot$10$^{-7}$&$4\cdot10^{-5}$&-0.32$\pm$0.50&0.23, 3.3 & 7.&2800 \\
NE.,$\Lambda$&40.4&14.5
&3.2$\cdot 10^{-6}$&0.87$\cdot 10^{-4}$ &-0.3$^{+1.}_{-0.4}$&0.43, 
5.0&5.&3000\\
SF1,M      &29.0&7.2&0.015&0.043&-0.49$\pm$0.08 &2.8, 9.2 &33 &1800\\
SF1,$\Lambda$&29.4&6.7&0.023&0.050&-0.48$\pm$0.09 &2.9, 16.& 28 &2000 \\
SF2,M      &30.2&6.8&0.018&0.031&-0.44$\pm$0.11 &2.9, 15.&25& 2700 \\
SF2,$\Lambda$&30.3&6.0&0.021&0.028&-0.43$\pm$0.12 &3.4, 21.&19& 3000 \\
SF3,M      &30.9&7.2&0.014& 0.20&-0.38$\pm$0.12&2.9, 11.&21&4400\\
SF3,$\Lambda$&30.7&6.55&0.014& 0.17&-0.34$^{+0.17}_{-0.14}$&4.2, 21.&16& 4500 
\\
\enddata
\label{fits}
\tablecomments{
The LF is parametrized as a truncated broken 
power law with 4 free parameters (the BPL model in \S 3.3). 
The number of degrees of freedom is 25. 
Many details are given in the Note of Table 1.
$\alpha$ is the slope of the
power law part of the LF below the break. 
The errors correspond to a $\Delta \chi^2 = 4$ confidence interval. 
$\Delta I_0$ is the confidence interval ($\Delta \chi^2 = 4$) 
of the break in the LF,
$I_0$, in photons s$^{-1}$ cm$^{-2}$  at $z=1$, 
Introducing a break in a single power law gives the $\chi^2$
reduction, $\Delta \chi^2 $, in the next column.
$N_{\rm tot}$ is the lower limit of the total GRB rate per year.
}
\end{deluxetable}

\centerline{\epsfxsize=9.5cm\epsfysize=10cm {\epsfbox{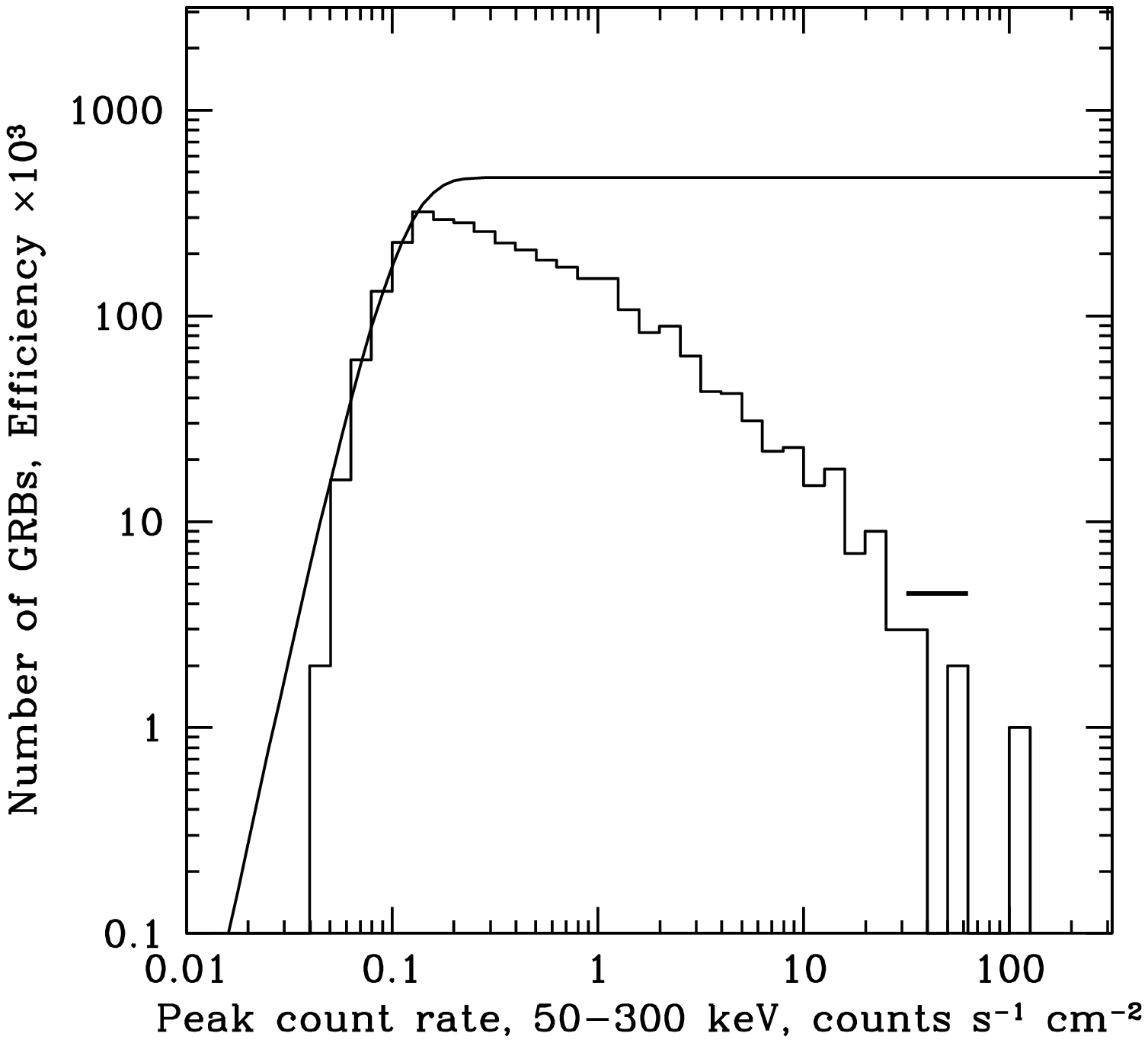}} }
\figcaption{
Fitted data. The histogram is the raw count rate distribution
of the sample of 3255 long GRBs found in the continuous  
BATSE records. The solid curve is the efficiency 
function $E(c_e)$ in equation (1)
measured using the test burst method. The discrepancy at the threshold 
results from the brightness bias (see eq. 1) and is removed using the 
efficiency matrix.
The thick horizontal bar on the right tail of the 
distribution shows the count rate contribution of the three intrinsically
brightest events  with measured redshifts if they were emitting at $z$=1. The
number of these  events
is renormalized to the 9.1 years of the BATSE observations. 
Their original number is 3 over $\sim$ 2 years.
}
\bigskip
\centerline{\epsfxsize=9.5cm\epsfysize=10cm {\epsfbox{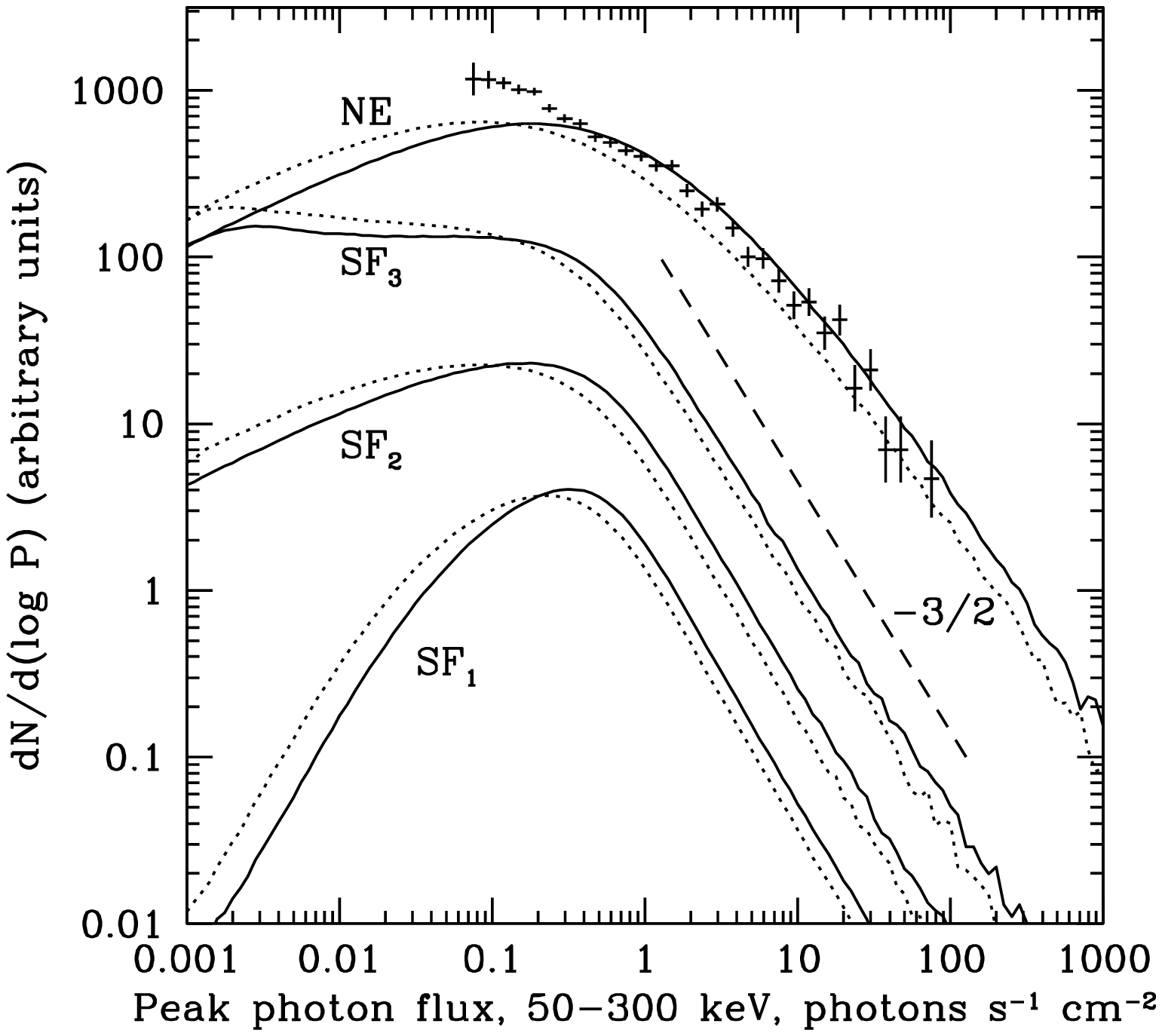}} }
\figcaption{
Standard candle log $N$ - log $P$ distributions for different models.
The standard candle brightness corresponds to a
photon peak flux of 1 photon s$^{-1}$ cm$^{-2}$
at $z$ = 1. From top to bottom: no evolution (NE), SF$_3$ (eq. 6),
SF$_2$ (eq. 5),  SF$_1$ (eq. 4). 
Solid curves - ($\Omega_{\rm M}, \Omega_\Lambda) = (1,0)$,
dotted curves - ($\Omega_{\rm M}, \Omega_\Lambda) = (0.3, 0.7)$.
The crosses represent the observed log $N$ - log $P$ distribution
of 3255 long BATSE GRBs.   
}
\bigskip

\centerline{\epsfxsize=9.5cm\epsfysize=10cm {\epsfbox{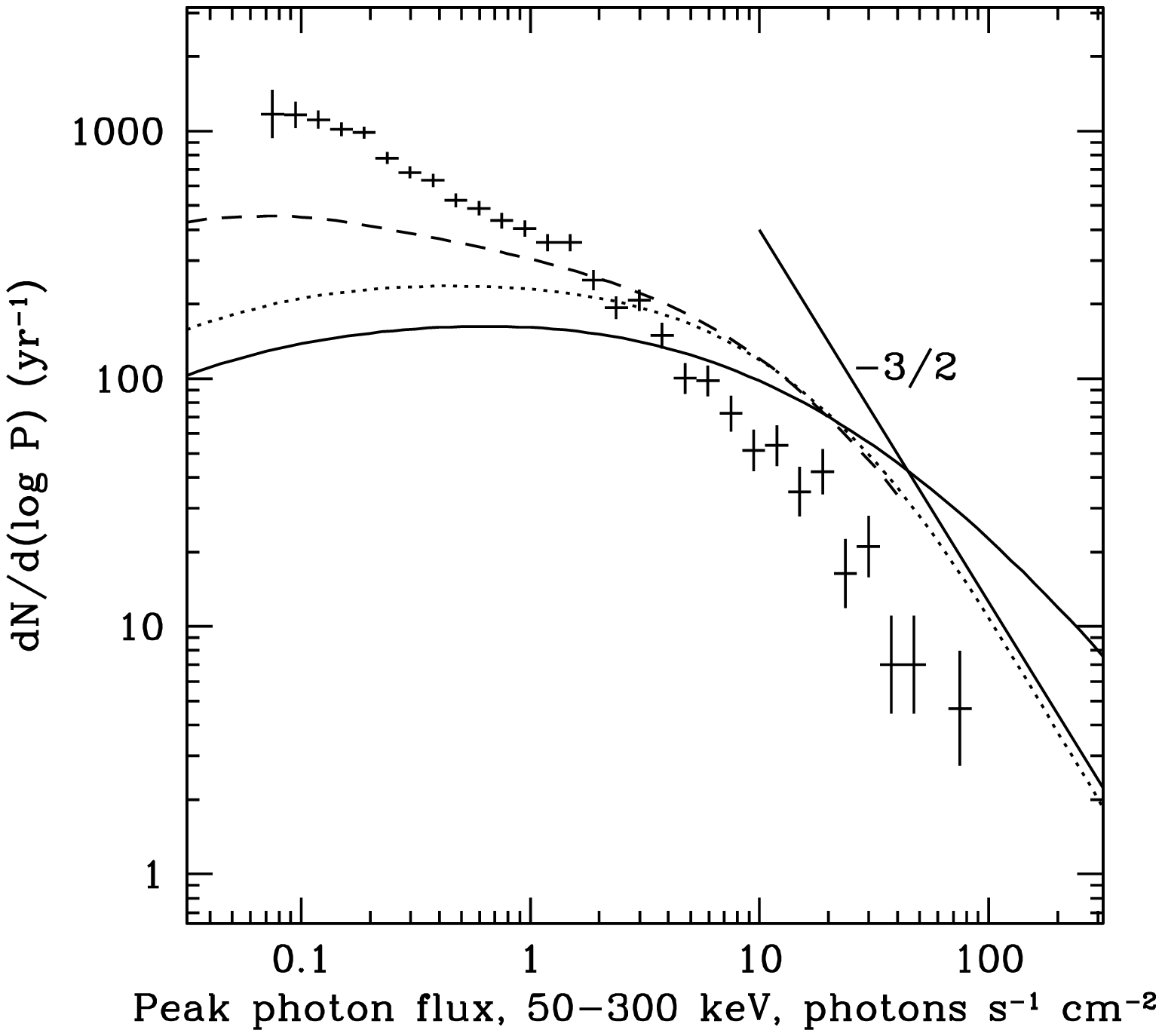}} }
\figcaption{
Test for the luminosity distribution inferred from the sample of 17 eve
nts
with known redshifts and intrinsic luminosities. Log $N$ - log $P$
distributions obtained with the luminosity distribution of this sample are
compared to the  BATSE data (crosses) for different evolutionary models.
Solid curve: NE model; dotted curve: SF2 model; dashed curve: SF3 model. 
The striking disagreement which can not be compensated with a 
reasonable redshift distribution shows that one can not rely on redshift data 
in the whole luminosity range because of a strong luminosity dependent 
selection bias. The used cosmology was 
($\Omega_{\rm M}, \Omega_\Lambda) = (0.3, 0.7)$. 
 }
\bigskip
\centerline{\epsfxsize=9.5cm\epsfysize=10cm {\epsfbox{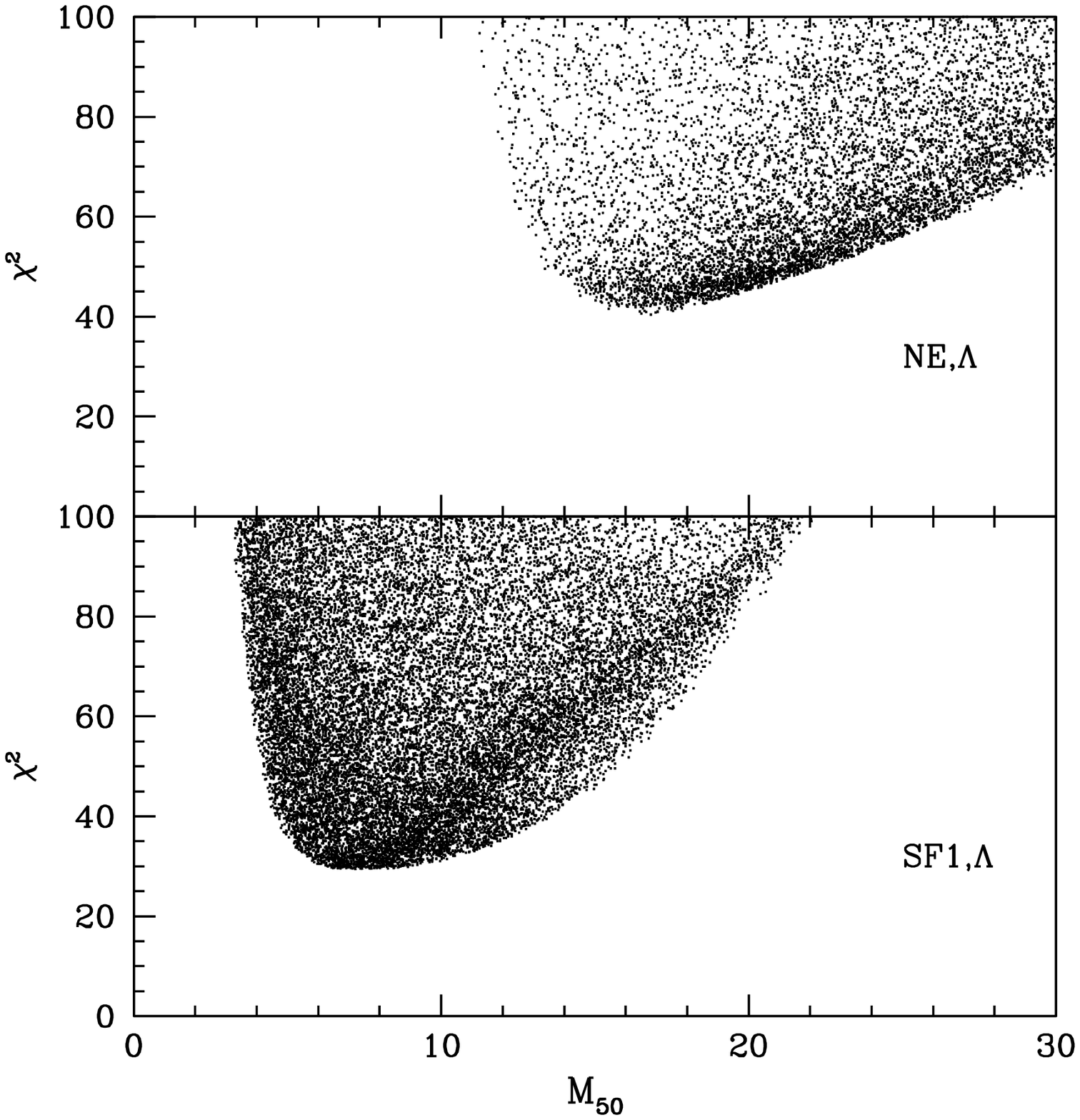}} }
\figcaption{
Fits of the bright tail of the log $N$ - log $P$ distribution using a  
broken power law LF.
The value of $\chi^2$  versus the predicted number of bright events
$M_{50}$ at $N_{40} > 3$ yr$^{-1}$. Actual value of $M_{50}$ is 3.
Parameters $I_1, I_0, \alpha$, and $\beta$ are random. The models are
NE,$\Lambda$ (upper panel) and  SF1,$\Lambda$ (lower panel). }
\bigskip

\centerline{\epsfxsize=9.5cm\epsfysize=10cm {\epsfbox{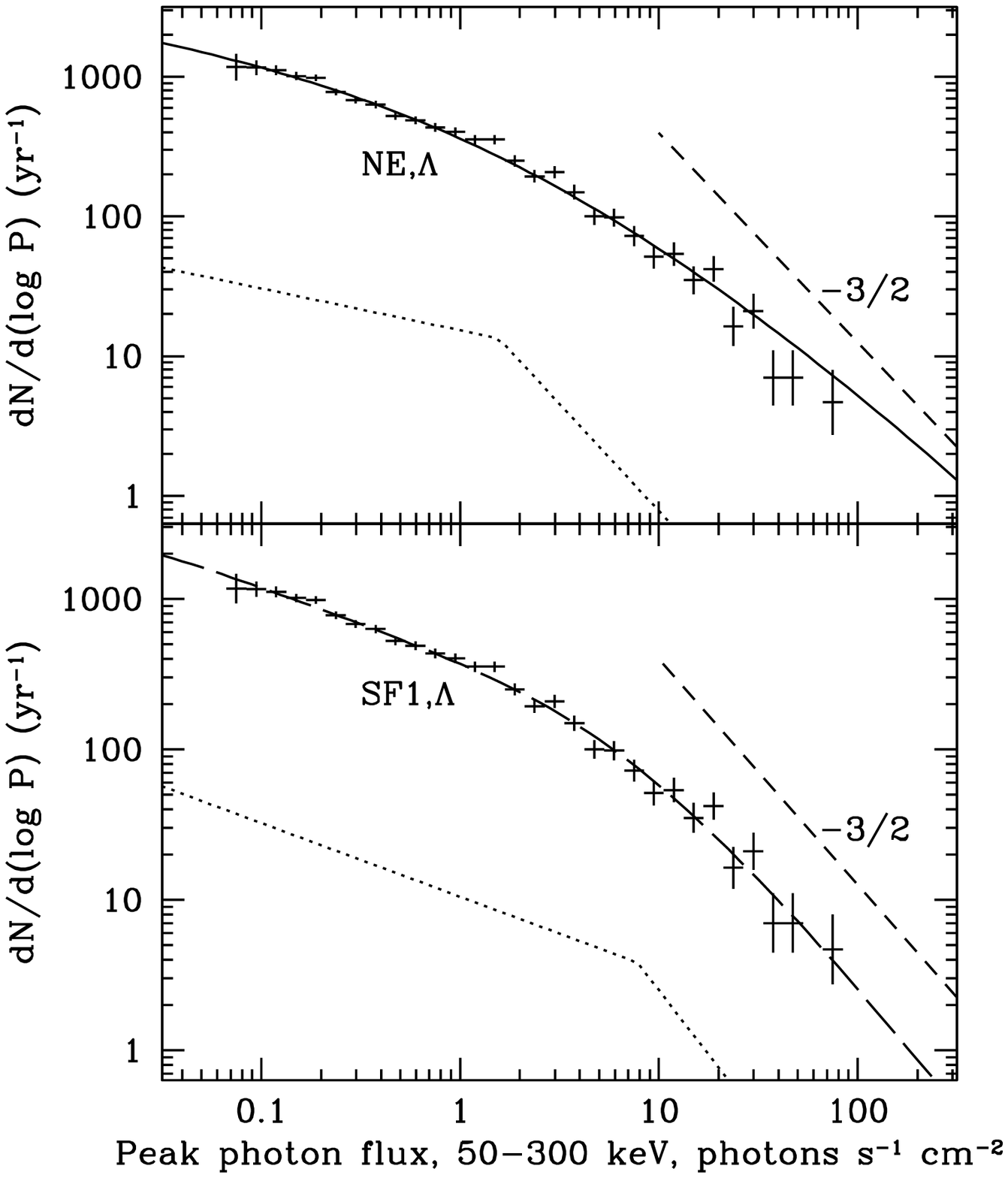}} }
\figcaption{
Best fits of data to models with no (upper panel, NE) and with SF1 
(lower panel) 
GRB source evolution.
Crosses are observed data points corrected using the efficiency matrix (1),
solid curves - the models, dotted curves - the model LF (a broken power law),
dashed line - the Euclidean -3/2 slope. The LF corresponds to a GRB luminosity
distance at $z=1$ with an arbitrary  normalization of the rate.  
The cosmological model has ($\Omega_{\rm M}, \Omega_\Lambda) = (0.3, 0.7)$.
For fitting parameters, see Table 2.
}
\bigskip
\centerline{\epsfxsize=9.5cm\epsfysize=10cm {\epsfbox{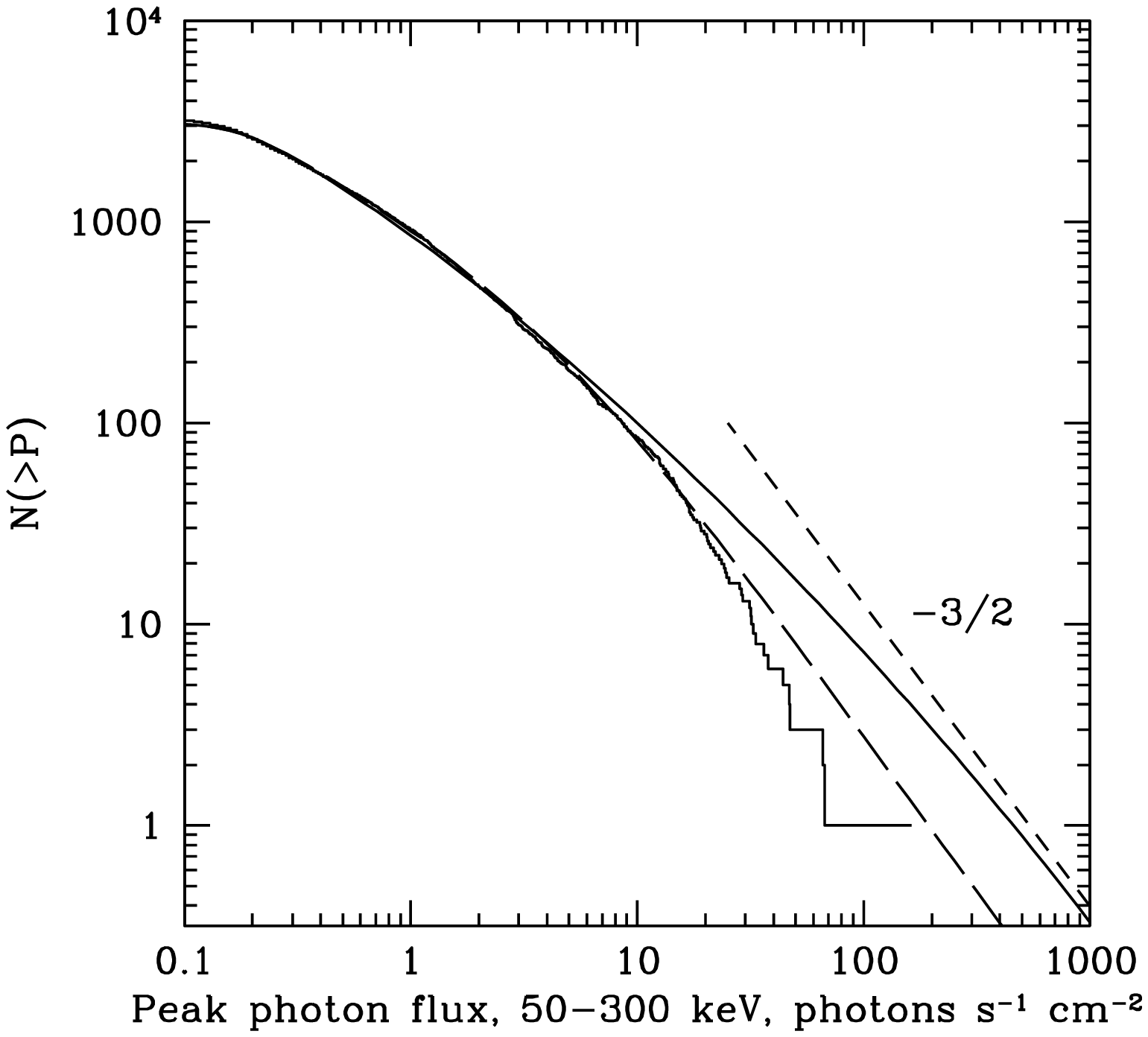}} }
\figcaption{
The same best fit functions as in Figure 5 but in integral form. 
Histogram - the raw peak count rate distribution in integral form
($N(>P)$ - the number of GRBs with peak flux  larger than $P$), solid curve -
the NE,$\Lambda$ model, dotted curve - the SF1,$\Lambda$ model, and dashed line 
- the
Euclidean distribution. The model curves are convolved with the efficiency 
matrix to
correspond to the raw data in this figure. }
\bigskip

\bigskip
\centerline{\epsfxsize=9.5cm\epsfysize=10cm {\epsfbox{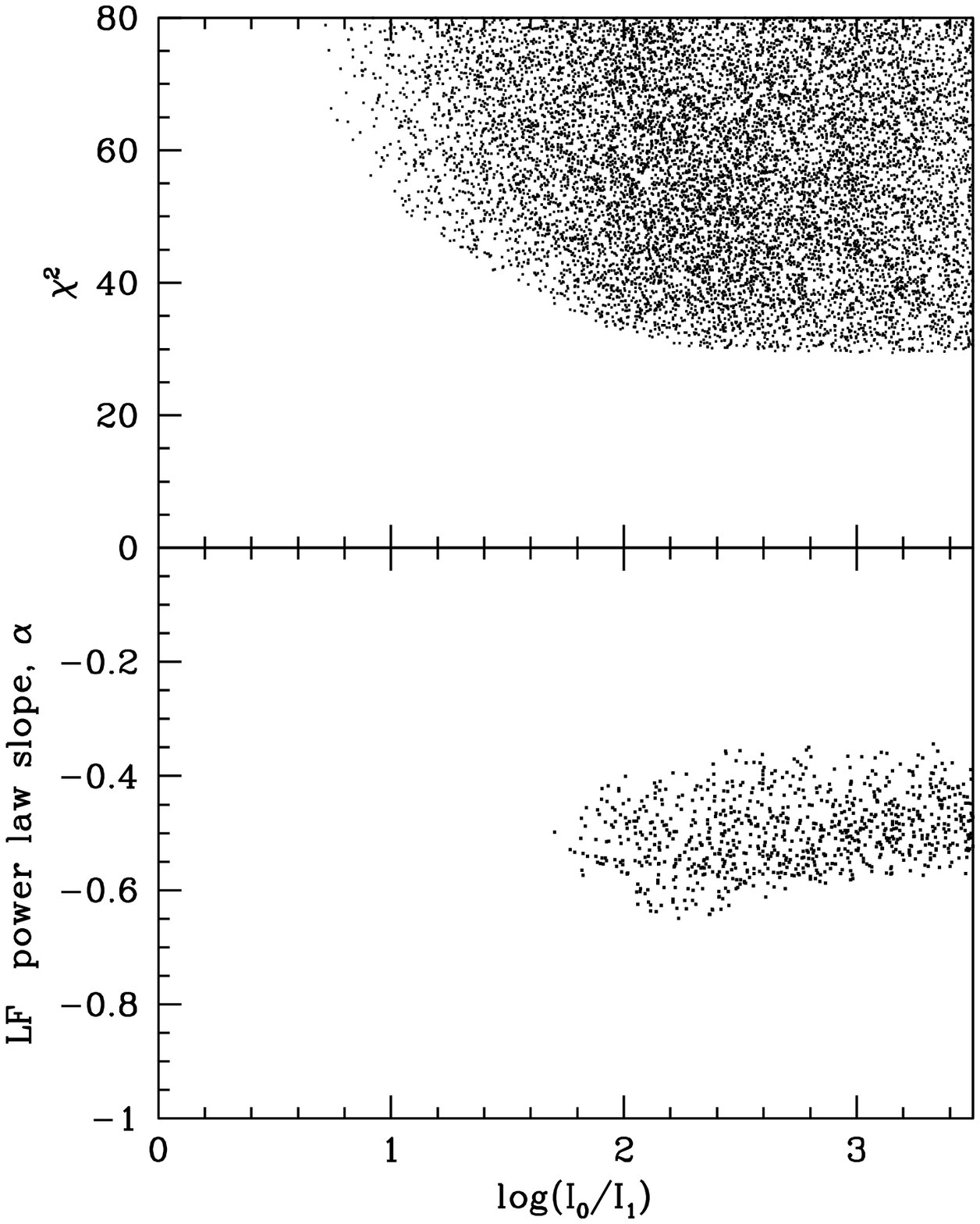}} }
\figcaption{ The confidence area for parameters of the power law
fragment  of the broken power law LF.
 The  model is SF1,$\Lambda$.  The confidence area corresponds to 
$\Delta \chi^2  < 6.17$, which formally corresponds to a 2 $\sigma$ 
confidence interval. However, see Notes of Table 1.
}
\bigskip
\centerline{\epsfxsize=9.5cm\epsfysize=10cm {\epsfbox{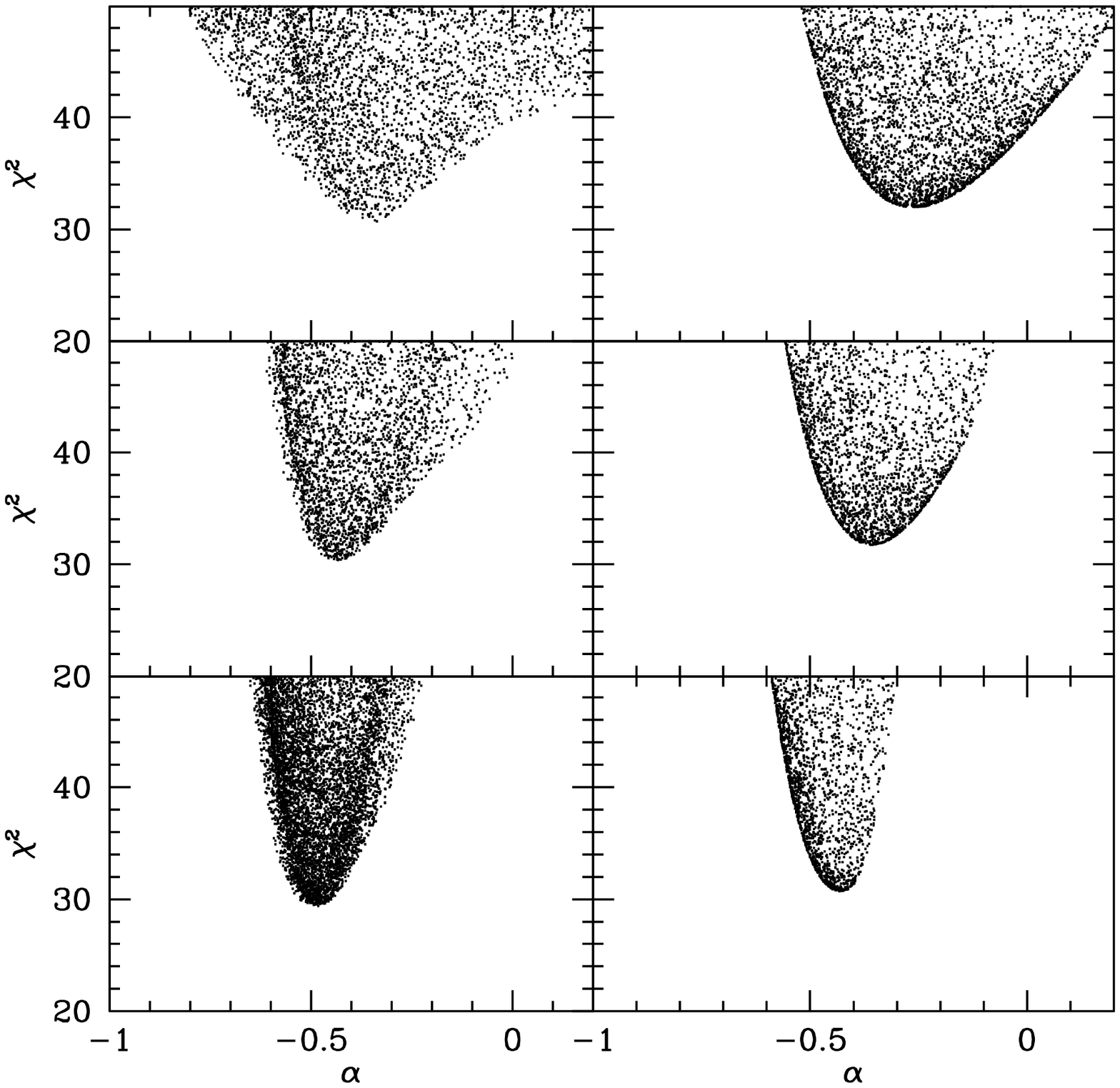}} }
\figcaption{ The profile of the $\chi^2$ minima for the power law
slope $\alpha$. Left panels: for a broken power law LF; right panels:
for the PLexp LF. {}From bottom to top: the SF1,$\Lambda$, SF2,$\Lambda$, and 
SF3,$\Lambda$ models.} 


\centerline{\epsfxsize=9.5cm\epsfysize=10cm {\epsfbox{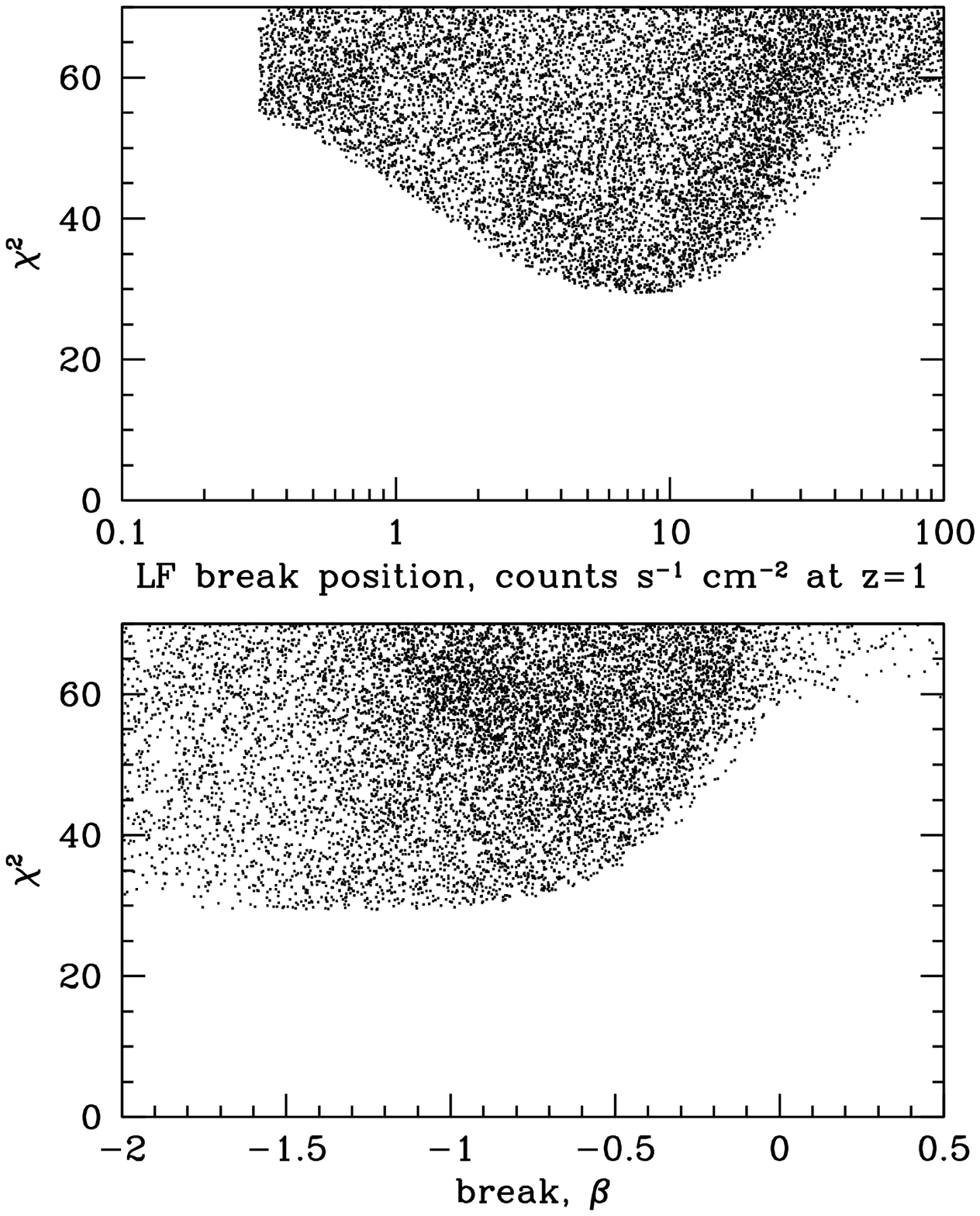}} }
\figcaption{ The characteristics  of the power law break in the broken
power law LF for the SF1,$\Lambda$ model. Upper panel: $\chi^2$ versus the 
position of the break, $I_0$, in the intrinsic peak brightness scale. 
Lower panel: 
$\chi^2$ versus the power law slope difference. Other parameters are random.   
}
\bigskip
\centerline{\epsfxsize=9.5cm\epsfysize=10cm {\epsfbox{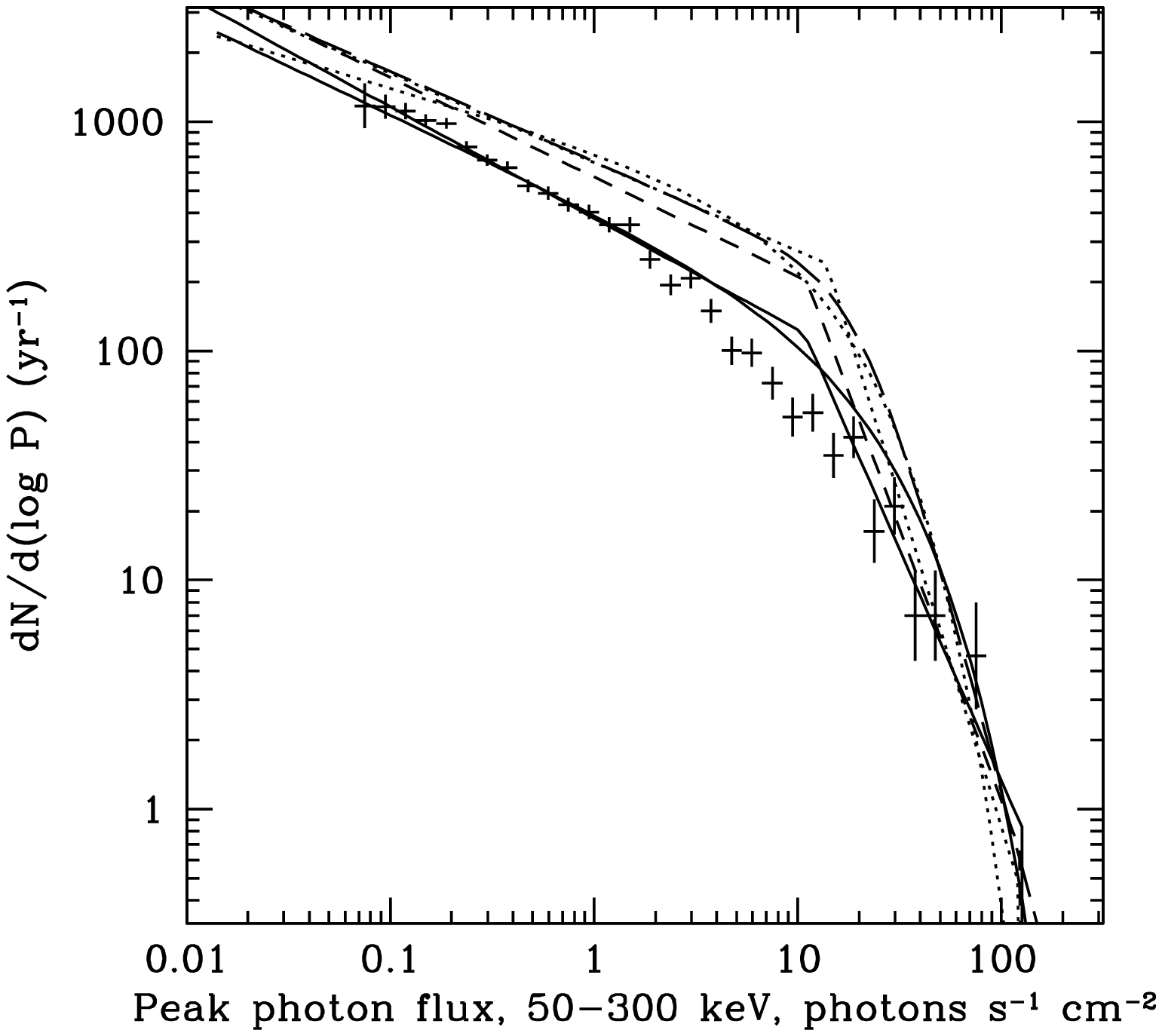}} }
\figcaption{
Best fit luminosity functions for different models.
Dotted curves:  a broken power law and the PLexp for the SF2,$\Lambda$ model;
dashed curve: the broken power law for the SF2,M model; 
dash-dotted curve: the smoothly broken power law (SBPL, see section 3.3)
for the SF2,$\Lambda$ model;
solid curves: a broken power law and the PLexp for the SF1,$\Lambda$ model.
The normalisation of the intrinsic brightness scale corresponds to the peak 
photon
flux being emitted at $z=1$. Crosses show real data points versus the 
{\it apparent} brightness, where the distance for each event is unknown.
Thus the ordinate for real data has a different meaning.
}
\bigskip


\begin{thebibliography}{}


\bibitem[Andersen et al. 2000]{z000131}
Andersen, M. I., et al. 2000, \aap, 364, L54

\bibitem[Atteia et al. 1999]{z000131}
Atteia, J.-L., Boer, M., \& Hurley, K. 2001,
in Proc. of the 19th Texas Symposium on Relativistic Astrophysics and Cosmology,
Ed. J. Paul, T. Montmerle, \& E. Aubourg (CEA Saclay), 4

\bibitem[Band et al. 1993]{band}
Band, D., et al. 1993, \apj, 413, 281

\bibitem[bloom et al 1999]{bloom}
Bloom, J. S., et al. 1999, Nature, 401, 453

\bibitem[Bulik 1999]{bulik}
Bulik, T. 1999, in ASP Conf. Series 190, Gamma-Ray Bursts: The First
Three Minutes, ed. J. Poutanen, \& R. Svensson (San Francisco: ASP), 219
(astro-ph/9911437)

\bibitem[Djorgovski et al. 1999d]{z990123}
Djorgovski, S. G., et al. 1999, GCN Circ. 251


\bibitem[Fishman et al. 1989]{BATSE}
Fishman, G. J., et al. 1989, in Proc. of the Gamma Ray Observatory
Science Workshop, ed. W. N. Johnson (Greenbelt: GSFC), 3


\bibitem[Hakkila et al. 1996]{hakkila}
Hakkila J. et al. 1996, \apj, 462, 125

\bibitem[Kommers et al. 2000]{kommers}
Kommers, J. M., Lewin, W. H. G., Kouveliotou, C., van Paradijs, J., 
Pendleton, G. N., Meegan, C. A., \& Fishman, G. J. 2000, \apj, 533, 696 

\bibitem[Krumholz, Thorsett, \& Harrison 1998]{krumholz}
Krumholz, M., Thorsett, S. E., \&Harrison, F. A. 1998, \apj, 506, L81

\bibitem[Lamb 1999]{lamb}
Lamb, D. Q. 1999, A\&AS, 138, 607 

\bibitem[Lloyd \& Petrosian 1999]
Lloyd, N. M. \& Petrosian, V. 1999, \apj, 511, 550 

\bibitem[Loredo \& Wasserman 1998]{lorwas}
Loredo, T. J., \& Wasserman, I. M. 1998, \apj, 502, 75


\bibitem[Lukash 2000]{luk}
Lukash, V. N. 2000 (astro-ph/0012012)

\bibitem[Panchenko 1999]{panchenko}
Panchenko, I. 1999, in ASP Conf. Series 190, Gamma-Ray Bursts: The First
Three Minutes, ed. J. Poutanen, \& R. Svensson (San Francisco: ASP), 271 
(astro-ph/9910450)

\bibitem[Panchenko 1999]{panchenko}
Panchenko I. E., Lipunov V. M., Postnov K. A., \& Prokhorov M. E. 1999,
   \aaps, 138, 517

\bibitem[Porciani \& Madau 2001]{porciani01}
Porciani, C., \& Madau, P. 2001, \apj, 548, 522

Portegeis Zwart, S. F., \& Yungelson, L. R. 1998, \aap, 332, 173

\bibitem[Schmidt 1999]{schmidt}
Schmidt, M. 1999, \aaps, 138, 409

\bibitem[Schmidt 1999]{schmidt}
Schmidt, M. 2000 (astro-ph/0001121)

\bibitem[Stern et al. 2000]{stern00}
Stern, B. E., Tikhomirova, Ya., Stepanov, M., Kompaneets, D., Berezhnoy, A.,
\& Svensson, R. 2000, \apj, 540, L21

\bibitem[Stern et al. 2001]{stern01}
Stern, B. E., Tikhomirova, Ya., Kompaneets, D., Svensson, R., \& Poutanen, J.
2001, \apj, 562, in press (astro-ph/0009447)

\bibitem[Totani 1999]{totani}
Totani, T. 1999, \apj, 511, 41

 

\bibitem[Paradijs et al., 2000]{paradijs}
van Paradijs, J., Kouveliotou, C., \& Wijers R. A. M. J. 2000, \araa, 38, 379

\bibitem[Wijers et al. 1998]{wj}
Wijers, R. A. M. J., Bloom, J. S., Bagla, J. S., \& Natarajan, P. 1998, 
\mnras, 294, L13

\bibitem[Vreeswijk et al. 1999b]{z991216}
Vreeswijk, P. M., et al. 1999, GCN Circ. 496


\end{thebibliography}
\end{document}